\documentclass[10pt]{article}
\usepackage[utf8]{inputenc}

\usepackage[T1]{fontenc}
\usepackage{graphics}
\usepackage{amsmath}
\usepackage{graphicx}
\usepackage{amsfonts}
\usepackage{amssymb}
\usepackage{newunicodechar}
\usepackage{xcolor}
\usepackage[margin=1in]{geometry}
\usepackage{authblk}

\usepackage{here}

\usepackage{hyperref}
\hypersetup{
    colorlinks=true,
    urlcolor=blue
}

\date{\today}

\title{Living Capillary Bridges}

\author[1,a]{Tytti Kärki}
\author[1,a]{Senna Luntama}
\author[2,a]{Yasamin Modabber}
\author[1]{Saila Pönkä}
\author[2]{Gonca Erdemci-Tandogan}
\author[2,3,4,5,6]{Mikko Karttunen}
\author[1]{Grégory Beaune}
\author[1,*]{Jaakko V. I. Timonen}
\affil[1]{Department of Applied Physics, Aalto University, 02150 Espoo, Finland}
\affil[2]{Department of Physics and Astronomy, Western University, London, Ontario, Canada}
\affil[3]{Department of Chemistry,  Western University, 1151 Richmond Street, London, Ontario, N6A\,5B7, Canada}
\affil[4]{European Laboratory for Learning and Intelligent Systems (ELLIS) Institute Finland, Maarintie 8, 02150 Espoo, Finland}
\affil[5]{Department of Technical Physics, University of Eastern Finland, P.O. Box 1627, FI-70211 Kuopio, Finland}
\affil[6] {Institut de Biologie Valrose, Université Côte d’Azur, Nice, France}
\affil[a]{Equal contribution: These authors contributed equally to this work}
\affil[*]{Corresponding Author}

\begin{document}

\maketitle

\section*{Abstract}
{\small

Biological tissues exhibit complex behaviors with their dynamics often resembling inert soft matter such as liquids, polymers, colloids, and liquid crystals. These analogies enable physics-based approaches for investigations of emergent behaviors in biological processes. A well-studied case is the 
spreading of cellular aggregates on solid surfaces, where they display dynamics similar to viscous droplets. \textit{In vivo}, however, cells and tissues are in a confined environment with varying geometries and mechanical properties to which they need to adapt. In this work, we compressed cellular aggregates between two solid surfaces and studied their dynamics using microscopy, and computer simulations. The confined cellular aggregates transitioned from compressed spheres into dynamic living capillary bridges exhibiting bridge thinning and a convex-to-concave meniscus curvature transition. We found that the stability of the bridge is determined by the interplay between cell growth and cell spreading on the confining surfaces. This interaction leads to bridge rupture at a critical length scale determined by the distance between the plates. The force distributions,  formation and stability regimes of the living capillary bridges were  characterized with full 3D computer simulations that included cell division, migration and growth dynamics, directly showing how mechanical principles govern the behavior of the living bridges; cellular aggregates display jamming and stiffening analogously to granular matter, and cell division along the long axis enhances thinning. Based on our results, we propose a new class of active soft matter behavior, where cellular aggregates exhibit liquid-like adaptation to confinement, but with self-organized rupturing driven by biological activity.

}

\newpage

\normalsize

\section*{Significance}

 The cellular level environment is crowded and structurally complex. Consequently, cells and tissues are constrained by complex boundaries, and they must adapt their mechanical properties to the locality. We use microscopy and \textit{in silico} modeling to investigate cellular response to confinement by constraining cellular aggregates between two plates. We observe the formation of living capillary bridges, bridge thinning, and a convex-to-concave meniscus curvature transition accompanied by jamming as demonstrated by the force distribution. The behavior manifests a new class of active soft matter behavior: Cellular aggregates exhibit liquid-like adaptation to confinement, and they can undergo self-organized rupturing driven by biological activity. The results help to understand cell invasion from tumors under physical stresses, and mechanobiology of self-organization under confinement.

\section*{Introduction}

Biological tissues are highly complex structures consisting of cells that interact with each other and their environment, leading to emergent collective behaviors. These behaviors include cellular mechanisms, such as cell division and migration, which give rise to processes such as embryogenesis, the maintenance of tissue homeostasis, and the onset of pathologies such as cancer \cite{Ruoslahti1996, zhu_principles_2020, mao_mechanical_2024}. Directly probing these processes \textit{in vivo} is often limited by tissue heterogeneity and restricted accessibility. To overcome these challenges, simplified cell, tissue and tumor models have been developed, ranging from 2D cell cultures to 3D structures such as cellular aggregates, which capture many
aspects of tissue organization and mechanics \cite{gonzalez-rodriguez_models, Velasco2020}. 

Cellular aggregates exhibit liquid-like behaviors, such as becoming spherical, fusion and viscous flows when placed on solid surfaces \cite{Holtfreter1943, steinberg_reconstruction_1963, stirbat_fine_2013, Foty2005, Douezan2011, Douezan2012, Beaune2014}. Additionally, cells grown in hollow geometries create smooth curved interfaces, resembling equilibrium liquid shapes \cite{rumpler_effect_2008, Bidan2013, Kychala2013, Kollmannsberger2018}. The liquid-like behavior of tissues has been attributed to surface tension and viscosity \cite{Holtfreter1943, steinberg_reconstruction_1963, manning_coaction_2010, stirbat_fine_2013, Foty2005}, which influence the shape and dynamics of tissues, similarly to inert liquids. Single cells can exhibit viscoelastic responses, which stem from their internal biopolymer networks \cite{KASZA2007101}, analogous to non-living polymer-based viscoelastic materials, \cite{Cross2012}, which contribute to the emergent behaviors and responses observed in cells. These mechanical properties in living cells and tissues, although their molecular origins are still debated, are crucial for understanding the fundamental tissue behavior, such as embryo compaction \cite{maitre_pulsatile_2015, firmin_mechanics_2024}, and cell migration driving cancer invasion \cite{Beaune2018, boot_spheroid_2021}. However, in contrast to inert soft matter, tissue consistency and mechanical properties can change in time, making them considerably more challenging to model than inert fluids. 

A wide range of models have been applied to tissues and cellular aggregates. Continuum approaches treat cellular aggregates as viscous or viscoelastic drops \cite{Douezan2011, Douezan2012, Beaune2014} or using a phase-field~\cite{Palmieri2015-oo}, while vertex and network models \cite{Nagai01072001, FARHADIFAR20072095, bi_density-independent_2015} capture cell-cell interfaces, rearrangements, and exhibit jamming transitions between fluid- and solid-like states \cite{mierke_ecm_2025}. Moreover, intrinsic cellular activities, such as divisions, apoptosis and migration, can be incorporated into these models as active stresses or self-propulsion at the single-cell level \cite{prost_active_2015}. Another family of models includes cell-based biomechanical 3D models such as Interacting  Active Surfaces (IAS) \cite{Torres-Sanchez2022-io}, SimuCell3D \cite{Runser2024-xb}, and \textsc{CellSim3D} \cite{Madhikar2018-fp}. From the continuum approaches, one particularly well-studied soft matter analogy in tissue dynamics is the collective migration of cells in spreading cellular aggregates, which resembles viscous droplet spreading \cite{Douezan2011, Douezan2012, Beaune2014}. The spreading of a cellular aggregates on solid surfaces has been described by a surface-energy-based model, originally developed for inert liquids \cite{Douezan2011, Douezan2012, Beaune2014}. In the model, the dynamics are primarily determined by the cell-cell adhesion energy ($W_\mathrm{CC}$) and cell-substrate adhesion energy ($W_\mathrm{CS}$). The spreading parameter $S = W_\mathrm{CC} - W_\mathrm{CS}$ determines how the tissue behaves on the substrate. When $S>0$, the cellular aggregate spreads completely forming a cellular monolayer, similarly as completely wetting liquids spread as a thin liquid film. 

Given the fluid-like behavior observed in cellular aggregates spreading on solid surfaces, here we explore whether cellular aggregates confined between two plates would create capillary bridges. Such living capillary bridges would demonstrate additional analogy between liquids and tissues, as capillary bridge formation is a characteristic of inert liquids. Previous work has created tissue capillary bridges by growing cells on pre-existing curved scaffolds \cite{Ehrig2019}, but we are interested in tissue adaptation and self-organization into such structures. Since capillary bridge shapes depend on volume, adhesion and local pressures, living capillary bridges provide an effective system to study how cellular activities shape bridging tissues by regulating cell number density, interactions and tensions. Additionally, liquid capillary bridges are known to exhibit instabilities \cite{Maeba2003}, which could also emerge in living tissues. In this paper, we aim to create living capillary bridges, characterize their dynamics, and understand the mechanical properties of such bridges.

\begin{figure}[htb]
    \centering
    \includegraphics[]{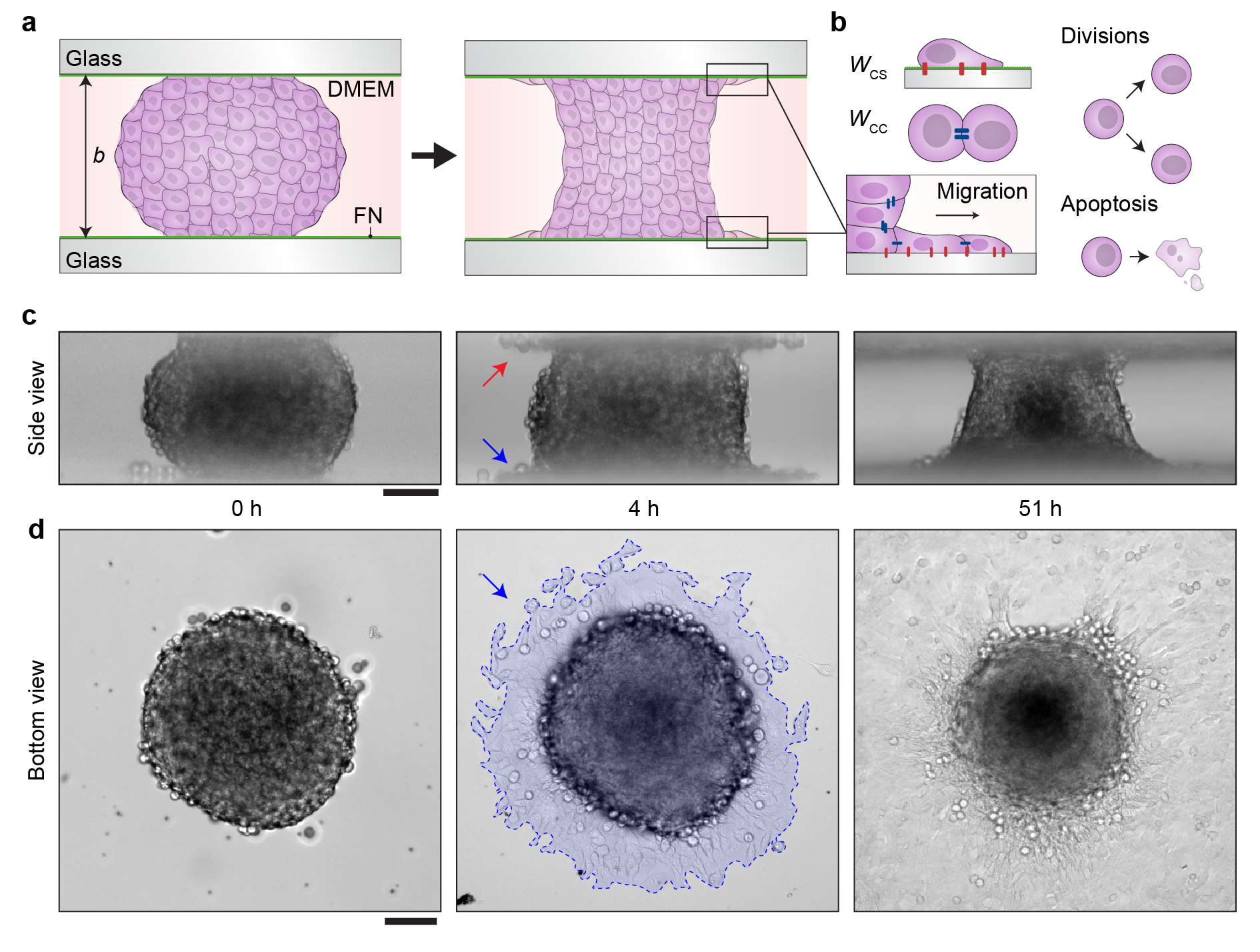}
    \caption{\textbf{The concept and geometry of living capillary bridges}. \textbf{a)} A scheme of a cellular aggregate confined between two fibronectin (FN) coated glass slides separated by distance $b$, spreading symmetrically on top and bottom surfaces. The aggregate is surrounded by cell culture medium (DMEM).
    \textbf{b)} Schemes of different active processes in the system. The cell-substrate and cell-cell adhesion energies (\textit{W}\textsubscript{CS}, \textit{W}\textsubscript{CC}) are based on cell adhesion molecules, from which we highlight integrins (red) and cadherins (blue). 
    \textbf{c,d)} Experimental time series of confined cellular aggregate from side and bottom view  ($R_0$ = 180\,µm, $C$ = 0.3). Red and blue arrows indicate spreading cell monolayer on the top and bottom surfaces, respectively. Scale bars are 100\,µm.} 
    \label{fig:1}
\end{figure}

\section*{Results and Discussion}

\subsection*{Preparation of living capillary bridges }

We create living capillary bridges of mammalian cells by confining cellular aggregates between two glass plates separated by distance $b$ (Figure~\ref{fig:1}a). We used cellular aggregates made of mouse murine sarcoma 180 (S180) cells \cite{Mege1988}, which are widely used in spreading experiments due to their importance in studying cancer invasion \cite{Douezan2011, gonzalez-rodriguez_detachment_2013, Beaune2014, Beaune2017}. The aggregates, with initial radii $R_0$ ranging from 50 to 200\,µm, were confined in a custom chamber consisting of two identical fixed glass slides with tuneable compression ratio $C= 1-b/2R_0$ (Figure~S1). The compression ratio was set to $C\approx 0.3$--0.4 ($b\approx$ 80--250 µm) for secure constriction of the aggregates without causing excessive pressure or deformation that might damage the cells. To promote cell adhesion and migration, we coated the glass slides with extracellular matrix glycoprotein fibronectin (FN) and maintained compression for up to four days at 37$^{\circ}$C and 5\% CO\textsubscript{2}. 

Based on previous research with S180 aggregates spreading on FN-coated glass \cite{Douezan2011, Douezan2012, Beaune2014}, we anticipate the confined cellular aggregates to spread on both top and bottom surfaces. The spreading dynamics depend on the adhesion energy between the cells themselves ($W_\mathrm{CC}$), and the cells and their substrate ($W_\mathrm{CS}$), in addition to cell number changes due to divisions and apoptosis  (Figure~\ref{fig:1}b). The role of gravity is expected to be negligible or minimal, since $b$ and $R_0$ are below the cellular aggregate capillary length $\lambda_c\approx \sqrt{\gamma/\Delta\rho g} \approx $ 4\,mm, where $\gamma$ is surface tension ($\gamma \approx 6$\,mN/m) \cite{Guevorkian2010, Yousafzai2022}, $\Delta\rho$ is the density difference between the cellular aggregate and cell culture medium ($\Delta\rho \approx$ 50\,kg/m\textsuperscript{3}, \cite{czerlinski_determination_1987}), and $g =$\,9.8 m/s\textsuperscript{2}. 

The confined cellular aggregates adhered and spread symmetrically on both the top and bottom surfaces. This leads to the creation of a living capillary bridge observable from side and bottom views (Figure~\ref{fig:1}c,d). The side views portray the meniscus transitioning from a convex to a concave shape within two days. This transformation occurred simultaneously with compaction and smoothening of the meniscus boundaries, a phenomenon observed across repeated experiments (Figure~S2). The formation and dynamics of the living capillary bridges resemble inert capillary bridges, such as those formed by viscoelastic polymer paste (Figure~S3). However, several differences highlight the complexity of living capillary bridges, such as altered volume and actively moving contact lines.

\subsection*{Formation and quasi-stability of large living capillary bridges}

\begin{figure}[H]
    \centering
    \vspace{9pt}
    \includegraphics[]{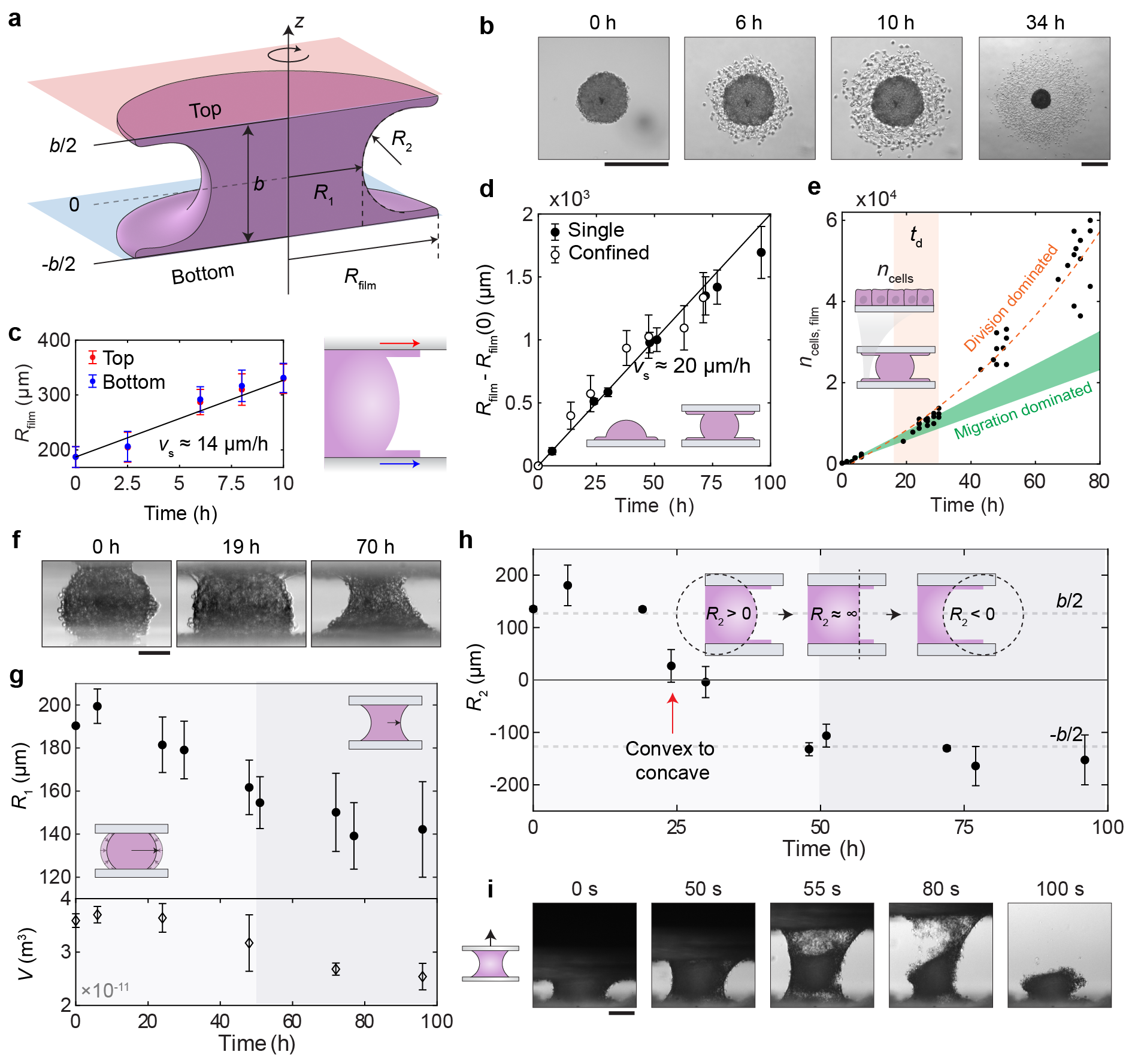}
    \caption{\textbf{Dynamics of large living capillary bridges}. \textbf{a}) Living capillary bridge geometry.
    \textbf{b)} Experimental bottom view time series of confined cellular aggregate ($R_0$ = 162\,µm; $C$ = 0.4). Scale bars are 500\,µm.
    \textbf{c)} The initial spreading of $R_\mathrm{film}$ as a function of time for confined cellular aggregate ($R_0$ = 180\,µm; $C$ = 0.3) on top and bottom plates. The linear fit approximates the spreading velocity $v_\mathrm{s}$.
    \textbf{d)} $R_\mathrm{film}$ as a function of time for confined cellular aggregate ($R_0$ = 180\,µm; $C$ = 0.3) and cellular aggregate spreading on one substrate ($R_0$ = 114\,µm). The linear fit approximates the spreading velocity $v_\mathrm{s}$.
    \textbf{e)} Approximated number of cells $n_\mathrm{cells, film}$ in a spreading film as a function of time. The sector (green area, upper and lower limits based on the range of division time $t_\mathrm{d}$) corresponds to an estimate for cellular film growth based solely on cell migration, and the exponential fit (dashed orange line) corresponds to film growth dominated by cell divisions.
    \textbf{f)} Time series of a living capillary bridge ($R_0$ = 180\,µm; $C$ = 0.3). Scale bar 100\,µm. 
    \textbf{g)} Living capillary bridge $R_\mathrm{1}$ and volume $V$ as a function of time ($R_0$ = 180\,µm; $C$ = 0.3). The volume estimate is based on the solid of revolution of the bridge outline.
    \textbf{h)} $R_2$ as a function of time ($R_0$ = 180\,µm; $C$ = 0.3).
    \textbf{i)} Time series of the forced rupture of a living capillary bridge ($R_0 = 160$\,µm, $C \approx$ 0.1). Scale bar 200\,µm. All error bars in (\textbf{c})-(\textbf{h}) indicate standard deviation from at least three experiments.}
    \label{fig:2}
\end{figure}

To investigate the formation dynamics of living capillary bridges, we first characterize large ($R_0 >$ 100\,µm) cellular aggregate dynamics by measuring the film radius $R_\mathrm{film}$, and the two orthogonal radii $R_1$ and $R_2$ at the bridge mid-plane ($z=0$)  as a function of time (Figure~\ref{fig:2}a). Starting from film dynamics (Figure~\ref{fig:2}b), the spreading velocity was initially $v_s\approx$ 14\,µm/h on both top and bottom surfaces (Movie~S1, Figure~\ref{fig:2}c), which increased to $v_s\approx$ 20\,µm/h after 10 hours (Figure~\ref{fig:2}d), comparable to values typically observed in non-confined, i.e. single surface geometry \cite{Douezan2011, Douezan2012, Beaune2014}. These results indicate that compression does not significantly affect the spreading dynamics. Since our experiments exceeded typical cellular aggregate spreading experiment timescales and cell doubling time ($t_d \approx$ 16-30 hours), we estimated the number of cells within the film as a function of time ($n_\mathrm{cells, film} \approx A_\mathrm{film}/A_\mathrm{cell}$, where $A_\mathrm{cell} \approx\pi R_\mathrm{cell}^2$ and $R_\mathrm{cell}\approx 7$\,µm) (Figure~\ref{fig:2}e). The results show a change in film growth dynamics around the cell doubling time. We assume that at short time scales (0--30 hours), spreading is mainly driven by the permeation of the cells from the aggregate to the film, and cell migration \cite{Douezan2011, Douezan2012, Beaune2014}. On longer time scales (30+ hours), cell division inside the spreading films begins to contribute as the cells in the film reorganize and adapt to the changing number of cells \cite{gonzalez-rodriguez_models, ranft_fluidization_2010}.

In contrast to the expanding cellular film, the living capillary bridge exhibited decrease in both $R_1$ and total volume, reaching a steady-state after approximately 50 hours  (Figure~\ref{fig:2}f,g). Therefore, we distinguish two regimes in the formation of living capillary bridges: a dynamic phase (from 0 to 50 hours) and a quasi-static phase (from 50 to 100 hours). The shift between these two phases was accompanied with convex-to-concave 
shape transition observable in the bridge radius $R_2$ (Figure~\ref{fig:2}h). However, the thinning of the bridge cannot be attributed solely to the shape transition in the meniscus, as a decrease in bridge volume was also observed (Figure~\ref{fig:2}g). Instead, alterations in the number of cells and/or compaction of the bridge contribute to its size reduction. This compaction at the bridge boundaries indicates formation of tension, which was strong enough to withstand a pulling force up to 2$b$, after which the bridge teared (Figure~\ref{fig:2}i). 

\begin{figure}[H]
    \centering
    \includegraphics[]{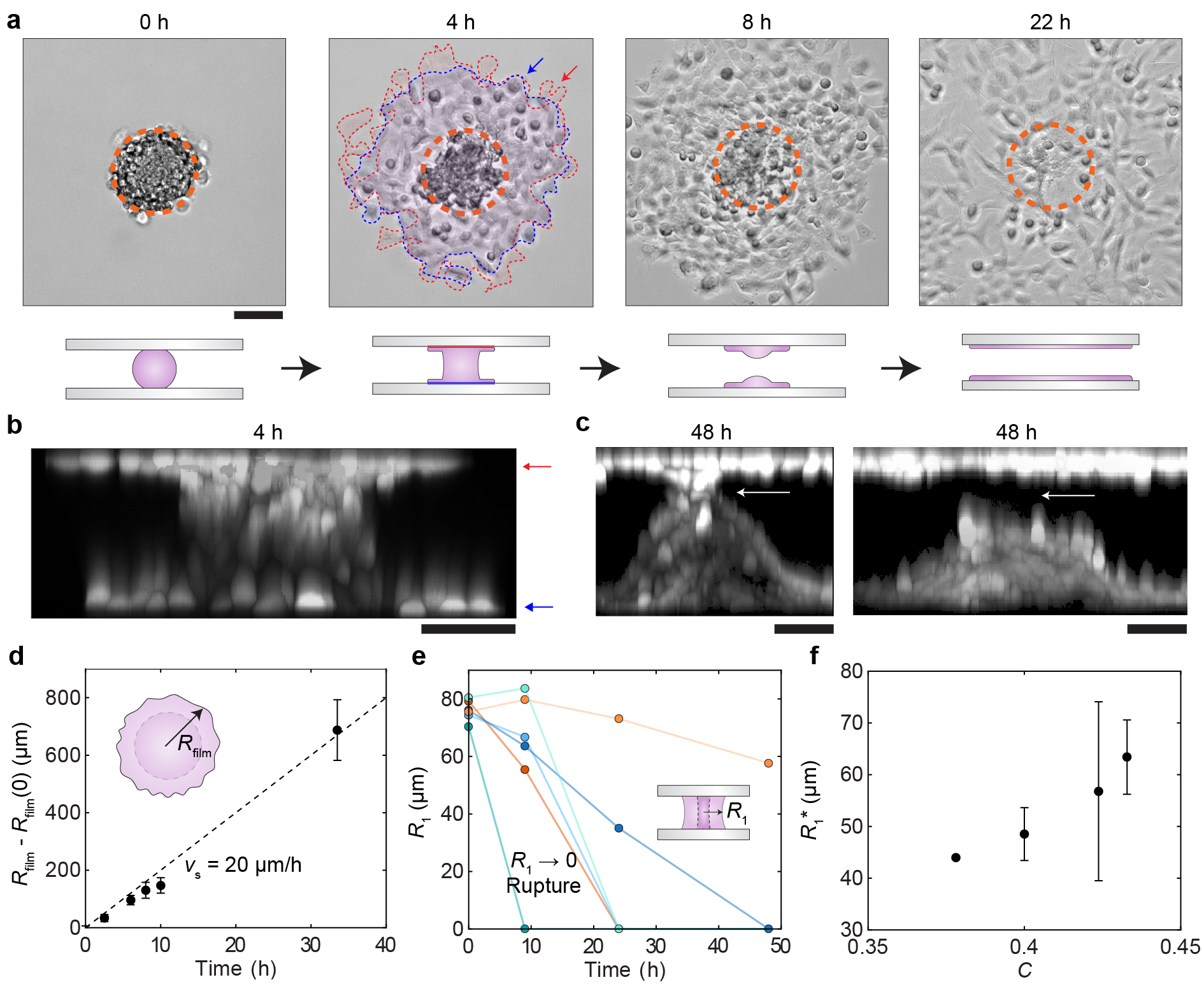}
    \caption{\textbf{Dynamics of small living capillary bridges. a)} Experimental bottom view time series of a rupturing living capillary bridge ($R_0$ = 81\,µm, $C$ = 0.4), and schematic of the rupturing living capillary bridge from side view. The upper (red) and lower (blue) spreading cellular films are visible by changing the focal plane. Scale bar 100\,µm.
    \textbf{b}) Confocal $z$-stack of small living capillary bridge after 4 hours of incubation ($R_0$ = 67\,µm; $C$ = 0.4). Scale bar 50 µm. \textbf{c}) Confocal $z$-stacks of small living capillary bridges after 48 hours of incubation ($R_0$ = 87\,µm; $C$ = 0.3). The white arrows indicate to rupturing regions. Scale bars are 50\,µm. \textbf{d}) $R_\mathrm{film}$ as a function of time ($R_0$ $\approx$ 40, 80 µm; $C \approx$ 0.4).  The linear fit approximates the spreading velocity $v_\mathrm{s}$. \textbf{e)} $R_1$ as a function of time from five individual living capillary bridges ($R_0$ = 82 µm; $C$ = 0.4). \textbf{f)} Critical bridge radius $R_{1}^*$ 2 hours before rupturing was observed $(R_0=$ 70 $\pm$ 20 µm). All error bars in (\textbf{d}) and (\textbf{f}) indicate standard deviation from at least three bridges.} 
    \label{fig:3}
\end{figure}

These findings highlight how cell reorganization and interactions can drive the formation of fluid-like, quasi-stable capillary bridges on the time scale of two days. Despite the continuous flux of cells from the bridge into the confining surfaces, the bridge maintained sufficient volume to counteract this outflow at least for four days. The observed volume reduction suggests, however, that initially smaller living capillary bridges might not be able to compensate for the outflux of cells even with cell growth. 

\subsection*{Instability of small bridges}

In our experiments, smaller cellular aggregates ($R_0 < 100$\,µm) exhibited similar spreading behavior on both the top and bottom surfaces as compared to the large confined cellular aggregates (Figure~\ref{fig:3}a). However, bottom view microscopy images showed that the darker region, corresponding to the bridge, disappeared after 1 day of incubation. We interpret this as the rupture of the living capillary bridge. Since side view imaging of small bridges is not feasible, we employed confocal microscopy to resolve the $z$-profiles. This confirmed the rupture of the bridges (Figure~\ref{fig:3}b,c). To ensure that the rupture was not caused by variations in the experimental conditions, we maintained a consistent compression ratio ($C \approx$ 0.3) and incubation conditions throughout all experiments. Additionally, reduced cell viability was ruled out as the cause since we observed positive staining with fluorescein diacetate (FDA), which permeates and is activated solely in viable cells (Figure~\ref{fig:3}b,c). Furthermore, the cells continued to spread and divide after rupturing indicating that they remained viable. Thus, the rupture of small living capillary bridges is a self-organized process resulting in the loss of bridge integrity.

The characterization of the small living capillary bridges showed similar spreading velocities ($v_s \approx$ 20\,µm/h) on both the top and the bottom surfaces, similar as observed in large bridges (Figure~\ref{fig:3}d). However, the small bridges exhibited faster decrease in the bridge radius often leading to bridge rupture (Figure~\ref{fig:3}e). Experiments with different initial aggregate sizes and compression ratios showed that rupturing occurred only when $R_0\le$ 100\,µm, suggesting that the size significantly influences bridge stability (Table~S1). Comparing the living capillary bridges to inert liquid capillary bridges of similar size and compression ratio indicates that bridges should remain stable, unless the plates are pulled apart, leading to Rayleigh-Plateau instability (if $b > R$, where $R$ is bridge perimeter) or if the volume reduces, for example, due to evaporation leading to the menisci collapse (fold bifurcation) \cite{slobozhanin_stability_1993}. To better understand the rupturing process of the living capillary ridges,  we estimated the critical mid-plane neck radius $R_{1}^*$ by measuring the smallest $R_1$ observed before bridge rupture ($\approx$ 2 hours before rupture). This analysis indicates that 
bridge cohesion is lost when $R_{1}^* \le 55 ~\pm$~11\,µm. Further,  comparison of $R_{1}^*$  from experiments with different initial radii ($R_0$) and compression ratios ($C$), indicates a correlation between $R_{1}^*$ and $C$ (Figure~\ref{fig:3}f), but not between $R_{1}^*$ and $R_0$. This could suggest that increasing the compression ratio increases the thinning velocity of the bridge.

\subsection*{Simple mathematical model reproducing quasi-stable and rupturing bridges}

\begin{figure}[H]
    \centering
    \includegraphics[]{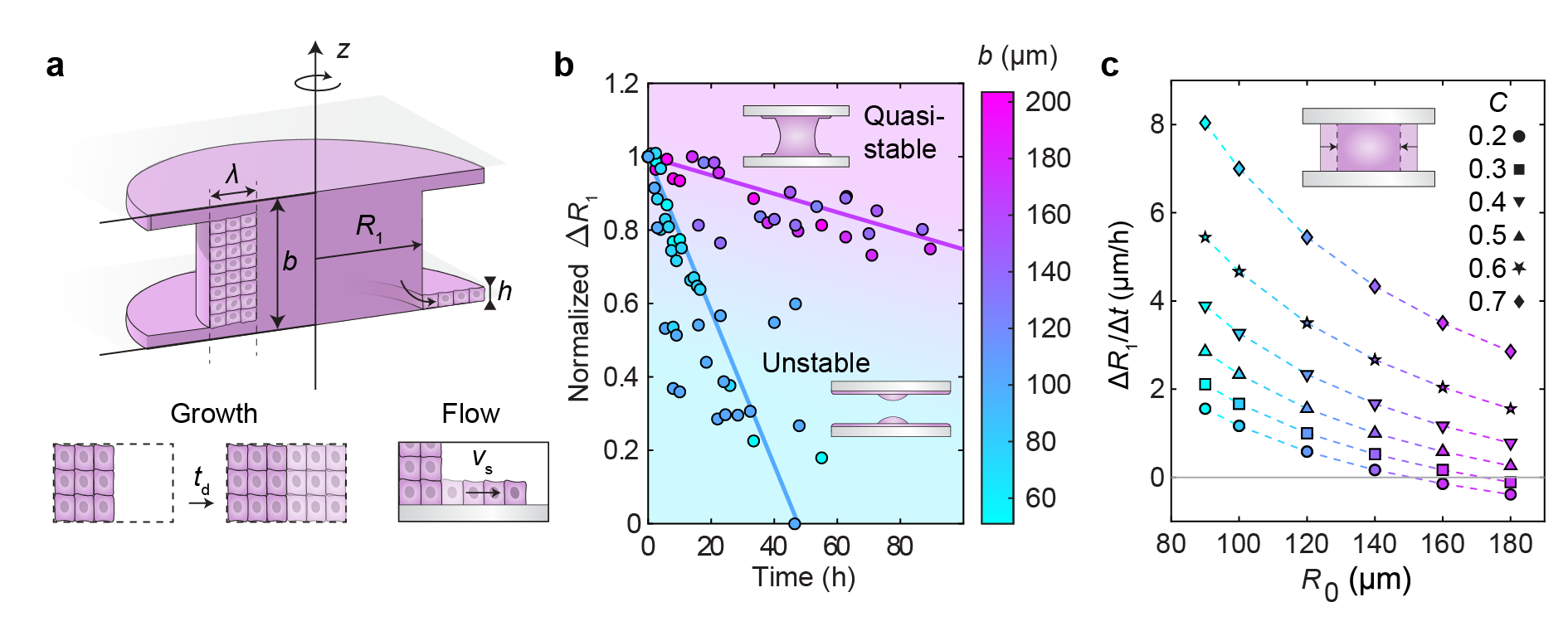}
    \caption{\textbf{Simple model for stable and unstable regimes of living capillary bridges. a)} A scheme of the cylindrical approximation of a living capillary bridge. The volume of the bridge increases due to cell growth in the proliferation layer with thickness $\lambda$ and rate determined by division time $t_d$, and decreases due to cell flow on the confining surfaces determined by velocity $v_m$. 
    \textbf{b)} Living capillary bridge $R_\mathrm{1}$ as a function of time for different compression lengths $b$ (compression ratio $C \approx$ 0.4) and initial aggregate sizes ($R_0 = $ 50--160\,µm). The colorbar indicates the compression length $b$. Linear fits correspond to model predictions with $b$ = 126\,µm (pink line) and 203\,µm (blue line) using $t_\mathrm{d} = 18$\,h.
    \textbf{c)} $R_1$ reduction velocity $\Delta R_1 / \Delta t$ as a function of initial confined aggregate radius $R_0$ according to the cylindrical approximation model. The markers correspond to different compression ratios $C$.}
    \label{fig:model}
\end{figure}

To build a minimal model for the dynamics of living capillary bridges, we approximate the volume of the living capillary bridge with a cylinder with radius $R_1$ and height $b$. The height is fixed, but the volume can increase due to cell divisions or decrease due to cell migration to the spreading films (Figure~\ref{fig:model}a):
\begin{equation}
    dV = dV_\mathrm{g}-dV_\mathrm{f}~,
    \label{eq:v_combined}
\end{equation}
where $dV$ is the volume change of the bridge, $dV_\mathrm{g}$ the volume growth and $dV_\mathrm{f}$ volume loss to the spreading film. We restrict the growth to a thin layer $\lambda$ on the surface of the living capillary bridge, where nutrients are expected to penetrate efficiently in the cells  \cite{Ackermann2021, carlsson_influence_1985}. Thus, the cell growth term is $dV_\mathrm{g} = \alpha V_\lambda dt$, where $\alpha = 1/t_\mathrm{d}$, $t_d$ is cell doubling time and $V_\lambda = 2 \pi R_1 b \lambda$, where $\lambda = 3\times 2R_\mathrm{cell}$, and for simplicity, we assume that $\lambda \ll R_0$ (Figure~\ref{fig:model}a). The volume loss term is given by $dV_\mathrm{f} = 2Q dt$, where $Q \propto h v_\mathrm{s} R_1$ is the volumetric flow rate, $h$ is the spreading cell layer thickness and $v_\mathrm{s}$ is the spreading velocity of the cells. The solution of Equation~\ref{eq:v_combined} yields
\begin{equation}
    R_1(t) = \left(\alpha \lambda - \frac{2 h v_\mathrm{s}}{b}\right) t + R_1(0) = k t + R_1(0)~,
    \label{eq:R1_model}
\end{equation}
where $R_1(0)\approx\sqrt{V_0/{\pi b}}$ (detailed derivation in Supplementary Note 1). 

Our model predicts bridge growth when $k > 0$, i.e. when $\alpha \lambda - 2 h v_\mathrm{s}/b> 0$, and a minimum critical confinement length $b_\mathrm{crit} = \frac{2 h v_\mathrm{s}}{\alpha \lambda}\approx $ 240\,µm for quasi-stable bridges. When $b < b_\mathrm{crit}$, the volume loss due to cell flow overpowers the cell growth in the bridge, resulting in the rupture of the bridge. The results from this model are in line with our experimental data, where we observed quasi-stable bridges when $b \approx 250$\,µm $> b_\mathrm{crit}$ and rupturing bridges when $b < b_\mathrm{crit}$ (Figure~\ref{fig:model}b). We can further explore the role of confinement and initial aggregate size $R_0$ for living capillary bridge stability by calculating the bridge thinning velocity $k = \Delta R_1/\Delta t$ (Figure~\ref{fig:model}c) which predicts that even with small compression ratios ($C = 0.2$), stable bridges $(\Delta R_1/\Delta t < 0)$ can be seen only with $R_0 \geq 150$\,µm. From this analysis we also see that increasing compression ratio increases thinning velocity, which is in line with our experiments (Figure~\ref{fig:3}f).

\subsection*{Computational cell dynamics modeling}

\begin{figure}[H]
    \centering
    \includegraphics[width=1\textwidth]{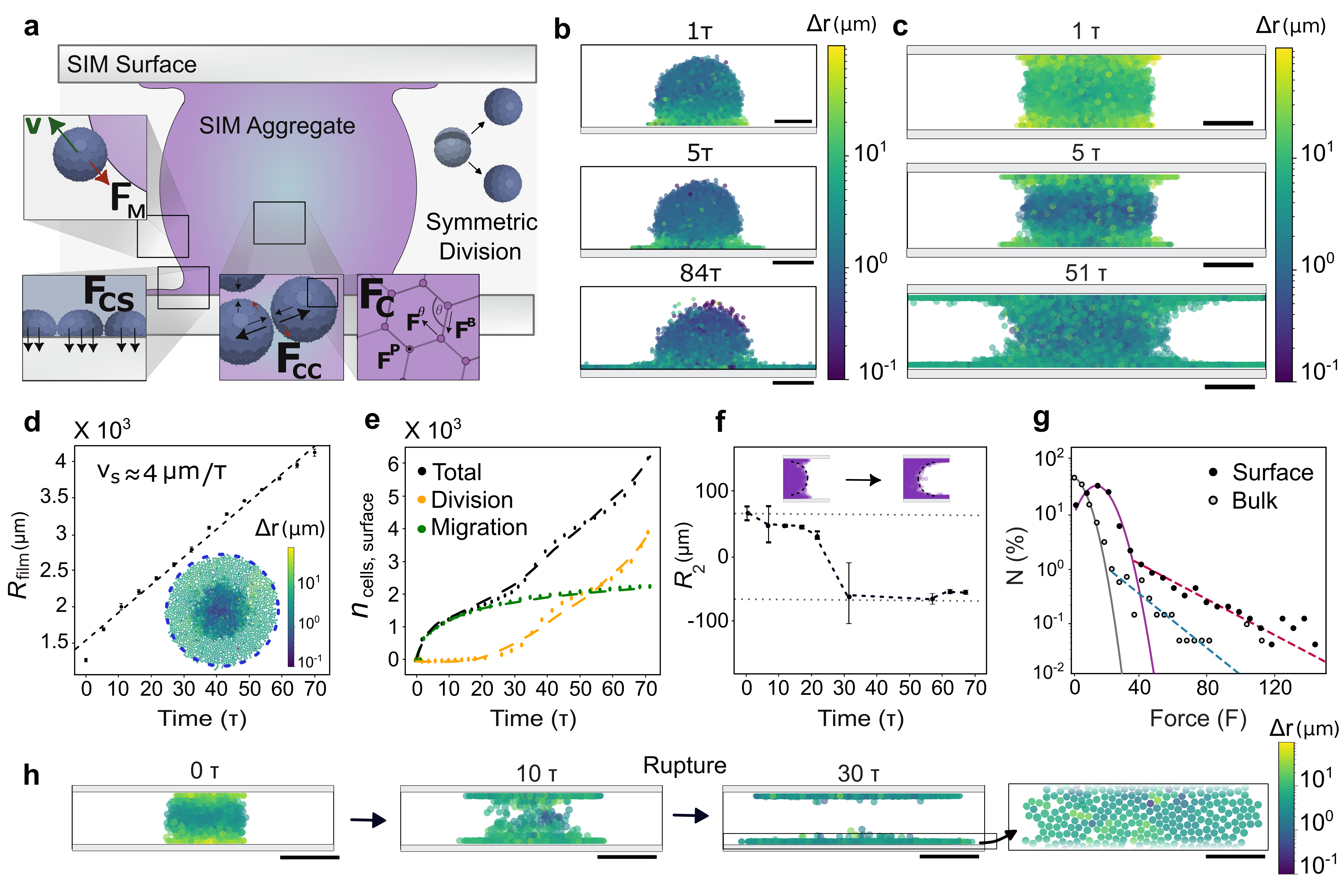}
    \caption{\textbf{Capillary bridge simulation model and dynamics.} 
    \textbf{a)} Schematic representation of the living capillary bridge simulation setup showing the main force components acting on the cells.
    \textbf{b)} Time series of simulated spreading aggregates  ($R_0$ = 100\,µm) on a single substrate at time points $t$ = 1 $\tau$, 5 $\tau$, and 84 $\tau$. $\tau$ represents the characteristic time scale of the simulation. The colorbar indicates cell displacement magnitude $\Delta r$ on a logarithmic scale, where $r$ corresponds to the position of the cell center of mass.
    \textbf{c)} Simulated living capillary bridge ($R_0$ = 100\,µm, $C$ = 0.3) at time points $t$ = 1 $\tau$, 5 $\tau$, and 51 $\tau$. The colorbar indicates cell displacement magnitude $\Delta r$ on a logarithmic scale.
    \textbf{d)} Simulated living capillary bridge film radius as a function of time ($R_0$ = 100\,µm, $C$ = 0.3). The linear fit (dashed black line) shows a spreading velocity of $v_s\approx$ 4\,µm/$\tau$. The insert plot shows bottom-view visualization of this capillary bridge. The colorbar indicates cell displacement magnitude $\Delta r$ on a logarithmic scale.
    \textbf{e)} Cell population dynamics on the simulated living capillary bridge spreading film. The green data points corresponds to migrated cells from the initial aggregate (at $t$ = 0), while the orange data points indicate the cell population of the younger cells resulting from cell divisions. The black data points are the total cell number.  The dashed lines highlight the overall trends. 
    \textbf{f)} Simulated living capillary bridge ($R_0$ = 100\,µm, $C$ = 0.3) meniscus curvature $R_2$ as a function of time. The dashed lines mark half confinement length ($\pm b$/2).
    \textbf{g)} Distribution of cell friction force at time $t$ = 31 $\tau$ , from simulated living capillary bridge ($R_0$ = 100\,µm, $C$ = 0.3) categorized by the spatial location. Gaussian fits: surface (purple, $A$ = 29.8, $\mu$ = 15.2, $\sigma$ = 8.5), bulk (gray, $A$ = 43.7, $\mu$ = 0.5, $\sigma$ = 7.3). Exponential fits: surface (red dashed line, $4.8e^{-F/{26.3}}$) , bulk (blue dashed line, $3.3e^{-F/{17.3}}$).
    \textbf{h)} Simulated capillary bridge ($R_0$ = 70\,µm, $C \approx$ 0.4) at time $t$ = 0 $\tau$ (intact), 10 $\tau$, and 30 $\tau$ (fully ruptured).The last figure depicts the bottom-view visualization of cells on the lower substrate at time $t$ = 30 $\tau$. The colorbar indicates cell displacement magnitude $\Delta r$ on a logarithmic scale
    }
    \label{fig4:schemes_and_validation}
\end{figure}

To further investigate the living capillary bridge formation and dynamics, we use \textsc{CellSim3D}~\cite{Madhikar2018-fp}, an open source coarse-grained model and software. In this model, cells are represented as C180 fullerene structures, that is, spherical polyhedra with 180 vertices connected by edges and faces. This geometric representation approximates the natural rounded shape of cells, observed in suspensions and tissues, while maintaining computational efficiency. This model was chosen for its ability to efficiently simulate large cell populations while providing a flexible parameter space that has been validated against experimentally measured tissue properties~\cite{madhikar_coarse-grained_2020}. The \textsc{CellSim3D} model incorporates intracellular interactions $\mathbf{F}_C$, intercellular interactions $\mathbf{F}_{CC}$, cell surface interactions $\mathbf{F}_{CS}$ , and cell medium interactions $\mathbf{F}_M$ (Figure~\ref{fig4:schemes_and_validation}a). This enables simulation of cellular processes such as growth, division, migration, cell-cell adhesion, apoptosis, and interactions with viscous media and rigid substrates. All divisions occur symmetrically, with each parent cell dividing into two daughter cells following Hertwig's rule, i.e., the mitotic spindle is aligned with the long axis, positioning the division plane perpendicular to it~\cite{hertwig_problem_1884, minc_influence_2011,moriarty_physical_2018}. Detailed descriptions of the model and the simulation parameters are provided in Supplementary Note 2.

First, to validate our approach, we simulated a classical system, i.e., cellular aggregate spreading on a single surface (Figure~\ref{fig4:schemes_and_validation}b). The simulations showed gradual spreading and flattening of the spherical cap as seen in experiments (Figure S3). Compressing an aggregate ($R_0 =$ 100\,µm) between substrates in the simulation resulted in the formation of a quasi-stable bridge (Movie~S2, Figure~\ref{fig4:schemes_and_validation}c) with dynamics similar to experiments (Figure~\ref{fig:1}c). In these simulations, long-axis division resulted in smaller bridge radii compared to randomly oriented division. The spreading film radius was measured over time, and a semi-linear growth, with a spreading velocity of $v_S\approx$ 4\,µm$/\tau$ (Figure~\ref{fig4:schemes_and_validation}d) was found ($\tau$ represents the characteristic time scale of the simulations). To reveal the motility patterns in the simulated capillary bridge, the 3D displacement magnitudes of individual cells were analyzed in a bottom-view cross-section of the aggregate, which revealed lower mobility in the interior compared to the outer radius. Next, leveraging the ability of simulations to separate and track individual cells based on their location, we measured the number of cells in the spreading film, and separated them into two populations: cells from the initial aggregate (at time $t=0$) and younger cells born from divisions (Figure~\ref{fig4:schemes_and_validation}e). Similar to the experiments, we observed that film growth was dominated by cell migration on shorter time scales and cell division on longer time scales). The meniscus curvature in simulated living capillary bridges underwent a gradual transition similar to that observed in the experiments (Figure~\ref{fig4:schemes_and_validation}f), confirming that the simulations capture the same underlying curvature dynamics driven by cell activity. 

To characterize the forces in the capillary bridges, we measured distributions of medium friction forces (viscous drag from the medium), categorizing the cells into those confined near the substrates (within one cell layer, thickness 14\,µm) and bulk cells in the interior.  Both populations followed Gaussian distributions at lower values (Figure~\ref{fig4:schemes_and_validation}g), but exhibited distinct exponential tails at higher forces, indicating that a subpopulation experiences larger drag forces compared to the rest. This effect was more pronounced in the population of cells in the surface layer, while bulk cells demonstrated lower overall friction and less mobility. The heterogeneous force distribution, combined with the immobile central regions identified in the displacement analysis (Movie~S3), signifies jamming in those areas \cite{OHern2001-lm,Madhikar2021-ko}. These localized jammed regions in the center of the aggregate solidify the bridges and contribute to their stability.  Simulations were also performed for smaller aggregates ($R_0 =$ 70\,µm) that formed unstable bridges, where the rupture occurred within  $\approx$14 simulation time steps (Movie~S4). This rupture, consistent with the experimental observations, demonstrates that small aggregates fail to develop the persistent low-mobility core necessary for stability.  Bottom view visualization after rupture revealed absence of low-mobility regions on the film (Movie~S5, Figure~\ref{fig4:schemes_and_validation}h), indicating that unlike in quasi-stable aggregates with a persistent low-mobility core, the cells in smaller living capillary bridges exhibit transient mobility patterns throughout, including at the central region.

\section*{Conclusions and Outlook}

Our experiments and simulations demonstrate that confined cellular aggregates can self-organize into living capillary bridges with size and shape influenced by collective cell migration and growth dynamics. Despite the complexities inherent to living tissues, the bridges exhibit characteristics similar to those of normal capillary bridges, such as spreading behavior determined by adhesion energies, quasi-static concave shapes and stability influenced by confinement length. However, the boundaries of living capillary bridges are complex and dynamic: The contact lines are not fixed but influenced by active cell migration and the meniscus undergoes smoothening driven by cell interactions.

These findings demonstrate how cellular interactions and reorganization can induce fluid-like behavior of tissues and tumors in confined geometries, which could be useful in understanding cellular reorganization in diverse environments and modeling their responses under tension. By investigating living capillary bridges formed by different cell types, with altered interactions, signaling cascades, or modified conditions such as varying ECM rigidity, we hope to gain significant insight into the mechanobiological principles governing self-supporting bridging tissues and the emergence of instabilities in tissues. Furthermore, since simulations effectively reproduce the dynamics of living capillary bridges, they serve as a valuable tool for identifying key parameters and guiding experimental design. Simulations also enable analyses that are difficult to perform experimentally, including quantifying force distribution patterns, tracking individual cell displacements to assess mobility and jamming, systematically controlling cell growth and division, and varying cell-cell interactions.

\section*{Acknowledgments and Funding sources}

T.K. and S.L. thank Sachin Rathod for helping with the rheometer, Kazusa Beppu for discussions, and Maja Vuckovac and Valtteri Turkki for their valuable help with force measurement trials, which provided useful insights during the development of this work.  
The Aalto University group was funded by Academy of Finland Center of Excellence Program (2022-2029) in Life-Inspired Hybrid Materials LIBER (346112) and Foundation for Aalto University Science and Technology. M.K. and G.E.-T. thank the Natural Sciences and Engineering Research Council of Canada (NSERC), M.K. thanks the Discovery and Canada Research Chairs Programs,  Initiative d'Excellence d'Université Côte d'Azur program, and Foundation PS, and G.E.-T. thanks the Western University Seed Grant for financial support.

\section*{Author Contributions}

Designed research: T.K., S.L., Y.M., G.E-T., M.K., G.B., J.V.I.T.; 
Performed research: T.K., S.L., Y.M., S.P.; 
Analyzed data: T.K., S.L., Y.M., G.E-T., M.K., G.B., J.V.I.T.; 
Wrote the paper: T.K., S.L., Y.M., G.E-T., M.K., G.B., J.V.I.T.

\section*{Competing Interest}

The authors declare no competing interests.

\section*{Methods}

\subsection*{Cell culturing}

S180 murine sarcoma cells were cultured as adherent monolayers in Dulbecco’s Modified Eagle Medium (DMEM, Gibco, 10566016) with 10\% (v/v) fetal bovine serum (FBS, Gibco, A5256701) and 1\% (v/v) Penicillin-Streptomycin (Gibco 15140122) at 37$^\circ$C and 5\% CO\textsubscript{2}. Cellular aggregates were prepared in hanging droplets as follows: a confluent adherent culture was subcultured 1:2 (v/v), and on the following day, the confluent culture was diluted into CO\textsubscript{2}-saturated DMEM with varying concentrations. Typical cell concentrations for small ($R_0$ $\approx$ 50\,µm) and large ($R_0$ $\approx$ 200\,µm) aggregates were $10 \times 10^3$\,cells/ml and $130 \times 10^3$\,cells/ml, respectively. The cell suspension was then pipetted in 15\,µl droplets on a Petri dish lid, which was inverted, and the Petri dish was filled with phosphate buffered saline (PBS) to avoid the evaporation of the hanging droplets. The hanging droplet cultures were incubated for 3 days, after which the cellular aggregates were ready for experimentation. 

\subsection*{Fibronectin coating}

The glass slides (VWR, 6141552) were plasma treated (Henniker HPT-100) for 3 minutes using 80\% power, followed by fibronectin (FN, Sigma-Aldrich, F1141) coating, where 100\,µl of FN-PBS solution (1:9, v/v) was pipetted on parafilm, on which the glass slide was placed. The incubation was 45 minutes at room temperature, after which the glass was washed with PBS three times. The coated glass slides were stored in PBS at 4$^\circ$C until use.

\subsection*{Confining cellular aggregates}

The confinement of cellular aggregates required two fixed solid plates with a gap smaller than the cellular aggregate. Two magnets were placed in a 100\,mm Petri dish. A FN-coated glass slide was placed on top of the magnets (Supermagnete, M-SEW-02, or Neomagnete Z-015-003-R), and spacers (PrecisionBrand plastic shim set, 77644905) were placed on the ends of the glass slides. A small amount of cell culture medium was added on the glass, and cellular aggregates were placed on the glass slide with a pipette. Another FN-coated glass slide was carefully placed on top of the cellular aggregates, and magnets were placed on top of the second glass slide. The Petri dish was filled with cell culture medium and incubated at 37$^\circ$C and 5\% CO\textsubscript{2}. For experiments done with side view imaging, the complete sample cells with confined cellular aggregates were placed in a rectangular plastic container ($94\times68\times29$\,mm, Semadeni 1729) instead of a Petri dish. The forced rupture of living capillary bridge was done in a rheometer (Anton Paar, \#MCR 702e Space) with FN-coated glass substrates prepared similarly as described for the confinement sample cells.

\subsection*{Fluorescence staining}

The culture medium was removed from the confined cellular aggregate sample Petri dish and replaced with Fluorescein diacetate (FDA, Sigma-Aldrich 343209) in PBS. Incubation was done for 5 minutes at room temperature before starting the imaging.

\subsection*{Microscopy}

The bottom view microscopy was done with Leica DMI6000 and Nikon Eclipse Ts2-FL with 5-20$\times$ objectives. The side view imaging was done with a custom setup where a Mitutoyo Video Microscopy Unit (VMU-H) was coupled with 5$\times$ objective (M Plan APO 5$\times$) and Ximea USB-camera (MC050CG-SY). The illumination was Thorlabs white LED coupled to a collimator and diffuser (MWWHLP1, SM2F32-A, DG20-120). The custom microscopy setup was built on Thorlabs sliding frame with mounts allowing xyz-translation of the light source, sample and imaging unit positions. Confocal microscopy was done with Zeiss Axio Examiner LSM710 with Zeiss 20$\times$/0.4 korr. objective. 

\subsection*{Computational modeling and simulations}

The cell dynamics were simulated using \textsc{CellSim3D}~\cite{Madhikar2018-fp}, a coarse-grained molecular dynamics model where cells are represented as 180 node fullerene structures. The forces acting on each node include intracellular components (node-node bonding $F^B$, angular constraints $F^{\theta}$, osmotic pressure $F^P$), intercellular attraction and repulsion ($F^A$, $F^R$), as well as cell-cell and cell-medium frictional interactions ($F^{F,e}$,$F^{F,m}$), cell-surface interactions ($F^W$) and stochastic noise ($F^{\eta}$). The simulations were initialized with single cells grown to form aggregates with about 10,000 cells (Figure S4). Spherical clusters of size $R_0$ were then extracted and compressed between planar substrates to create the capillary bridge configuration. A description of the initialization process is also included in the Supplementary Information. 

\subsection*{Data analysis}

The recorded image data and confocal stacks were processed using ImageJ. Data from the bottom and side view imaging were measured manually in ImageJ or using custom MATLAB scripts utilizing the Image Processing Toolbox. The radius of curvature was estimated by first processing the side view image data to extract the meniscus boundary as a set of points and then fitting a circle to the boundary points based on algebraic least squares fitting. The simulation data was visualized and analyzed using custom Python scripts. The film outer bounds were detected by applying convex hull analysis to cell positions. All other relevant calculations were performed using data extracted from annotations on the output images. The radius of curvature and film radius were calculated by fitting circles to the data using the least squares method.


%

\end{document}